\documentstyle[epsfig,preprint,aps,pra]{revtex}
\begin{document}
\tighten
\title{Low--energy $\eta d$ resonance}
\author{
        N. V. Shevchenko$^{1,3}$, V. B. Belyaev$^{1,2}$,
	S. A. Rakityansky$^{2}$, S. A. Sofianos$^{2}$, W.~Sandhas$^4$}
\address{
$^1$Joint Institute  for Nuclear Research,Dubna, 141980, Russia}
\address{$^2$Physics Dept., University of South Africa,
         P.O. Box 392, Pretoria 0003, South Africa}
\address{$^3$
Physics Dept., Irkutsk State University, Irkutsk 664003, Russia}
\address{$^4$
Physikalisches Institut, Universit\H{a}t Bonn, D-53115 Bonn, Germany}
\maketitle
\begin{abstract}
Elastic $\eta d$ scattering is considered  within the
Alt--Grassberger--Sandhas (AGS) formalism for various $\eta N$ input data. A
three--body resonant state is found close to the $\eta d$ threshold. This
resonance is sustained for different choices of the two--body $\eta N$
scattering length $a_{\eta N}$. The position of the resonance moves towards
the $\eta d$ threshold when ${\rm Re\,}a_{\eta N}$ is increased, and turns
into a quasi-bound state at ${\rm Re\,}a_{\eta N}=0.733$\,fm.\\

{PACS numbers:  25.80.-e, 21.45.+v, 25.10.+s} 
\end{abstract}
%%%%%%%%%%%%%%%%%%%%%%%%%%%%%% 
%\section{Introduction} 

One of the main questions of $\eta$--meson physics concerns the existence of
$\eta$--nuclei, a possibility predicted in a pioneering work by Haider and
Liu \cite{Hai86}.  In this context it is worth mentioning that according to
\cite{Rol96} the mean free path of $\eta$ mesons in a nuclear medium is about
2\,fm, {\em i.e.}, less than the size of a typical nucleus.  A necessary
condition for the existence of $\eta$--nuclei, hence, appears to be
satisfied. A further indication is the observation of pionic nuclei (deeply
bound pion states inside nuclei) \cite{Yam}, which were predicted
theoretically in \cite{Tok88}.

Recently, in a number of theoretical investigations based on quite different
approaches, such as the mean field approximation \cite{Tsu}, the optical
model \cite{Kul}, and few--body calculations \cite{Rak96,She98}, the idea of
$\eta$--nuclei was given a rather firm ground.  Experimentally, after the
discouraging attempt of Ref.~\cite{Gri88}, certain evidence for the existence
of $\eta$--nuclei was noticed by two groups \cite{Wil99,Sok}.  An additional
indication is given by the enhancement of near--threshold $\eta$ production
in the reaction $np\to d\eta$ as reported by the Uppsala group \cite{calen}.
This observation is most likely to be understood as the effect of a
near--threshold bound or resonant $\eta d$ state.  Moreover, it was suggested
\cite{nefk} that even quite exotic systems containing an $\eta$ meson and a
hyperon ($\eta$--hypernuclei) may exist.

Being of interest by itself, the existence of $\eta$--nuclei would also shed
new light on various fundamental problems of particle physics.  For example,
the study of $\eta$--nuclei would give a clue for understanding the possible
restoration of chiral symmetry in a nuclear medium, or its partial
restoration which may occur at normal densities $\sim 0.17\,{\rm fm}^{-3}$
\cite{Hat94,Bro96,kim}. A further aspect, which makes such systems
interesting for other fields of nuclear physics, is the modification of the
two--body $\eta N$ force by the nuclear medium.  Being enhanced due to the
resonant character of the $\eta N$ interaction, general properties of such
modifications should become particularly evident in this case.  Especially,
the structure of the $S_{11}$ resonance embedded in a nuclear medium may be
studied. This would exhibit certain details of the effective Lagrangian
models of such a resonance \cite{Mab}, and shed some light, for example, on
chiral models which suggest a reduction of its partial width for decaying
into the $\eta N$ channel \cite{kim}, or assume $S_{11}$ to be a quasi-bound
state of the $K$ meson and a $\Sigma$ hyperon \cite{Kei97}. The interesting
suggestion of considering $S_{11}$ as a joint manifestation of threshold
(casp) and resonance phenomena \cite{Huh97} could also be checked.

In a sense, the problem of choosing the most adequate model for describing
the $S_{11}$ resonance is similar to the usual difficulty of choosing the
``correct'' two--body potential.  A reliable judgment on such a choice can be
given when this potential (or the resonance) is used in a few--body system
where its off-shell properties are exposed.  This requires a treatment based
on rigorous few--body theory.  The Alt--Grassberger--Sandhas (AGS) equations
\cite{Alt67}, which are employed in this work to calculate the
$\eta$--deuteron elastic scattering amplitude, belong to this category.

The Faddeev--type coupling of these equations guarantees uniqueness of their
solutions.  Moreover, as equations for the elastic and rearrangement
operators $U_{\alpha\alpha}$ and $U_{\beta\alpha}$, respectively, they are
well--defined in momentum space, providing thus the desired scattering
amplitudes in a most direct and technically reliable manner.  The advantage
of working with coupled equations is not only suggested by questions of
uniqueness, but also by the relevance of rescattering effects.  Indeed, in
recent calculations of $\eta$--photoproduction from the deuteron it was found
that rescattering terms give a significant contribution to the corresponding
amplitude \cite{Rit99}.

%=====================================================================

Let the $\eta$ meson be denoted as particle 1 and the two nucleons as
particles 2 and 3.  With the momentum ${\bf q}_1$ of the meson relative to
the center of mass of the nucleons in the deuteron state $|\psi_d\rangle$,
the free $\eta d$ channel state is given by
\begin{equation}
\label{w1}
  |\psi_d;{\bf q}_1\rangle =|\psi_d\rangle|{\bf q}_1\rangle\ ,
\end{equation}
its normalization being chosen as  
\begin{equation}
\label{w2}
      \langle\psi_d;{\bf q}'_1|\psi_d;{\bf q}_1\rangle =
      \delta({\bf q}'_1-{\bf q}_1)\ .
\end{equation}
In terms of the AGS transition operator $U_{11}$ the $\eta d$ elastic
scattering amplitude is represented as
\begin{equation}
\label{eampl}
	  f({\bf q}'_1,{\bf q}_1)=-(2\pi)^2M_1
	  \langle\psi_d; {\bf q}'_1|U_{11}(z)|\psi_d;{\bf q}_1\rangle\ ,
\end{equation}
with the on--energy--shell conditions $|{\bf q}'_1|=|{\bf q}_1|$ and
$z=E_d+q_1^2/2M_1+i0$.  Here $M_1$ is the reduced mass of particle 1 and the
(23) subsystem, $1/M_1=1/m_1+1/(m_2+m_3)$.\\

The elastic transition operator $U_{11}$ satisfies the set of AGS equations
\begin{equation}
     U_{\beta\alpha}(z)=(1-\delta_{\beta\alpha})G_0^{-1}(z)+\sum^3_{\gamma=1}
     (1-\delta_{\beta\gamma})T_\gamma(z)G_0(z)U_{\gamma\alpha}(z)\ ,
\label{AGS}
\end{equation}
with $G_0(z)$ being the free resolvent (Green's operator) of the three
particles involved.  This set of equations couples all $3\times 3$ elastic
and rearrangement operators $U_{\alpha\alpha}$ and $U_{\beta\alpha}$.  Here
each of the subscripts runs through the values 1,2 and 3, indicating the
two--fragment partitions (1,23), (2,31) and (3,12) respectively.  Therefore,
$U_{11}$ describes the elastic transition 1(23)$\to$1(23), while $U_{21}$
represents the rearrangement process 1(23)$\to$2(13).

In accordance with Eq.(\ref{w1}), we denote by ${\bf q}_\alpha$ the momentum
of particle $\alpha$ relative to the center of mass of the ($\beta\gamma$)
subsystem, and by $M_\alpha$ the corresponding reduced mass.  The internal
momentum of this subsystem is denoted as ${\bf p}_\alpha$.  Conventionally,
this complementary notation is also used to label the two--body $T$--operator
$t_\alpha(z)$ of the ($\beta\gamma$) pair, for instance $t_1(z)=t_{NN}(z)$.
It should be noticed, however, that it is not this genuine two--body operator
which enters the AGS equations, but the operator
\begin{equation}
\label{shift}
         T_\alpha(z)=t_\alpha(z-{\hat {\bf q}_\alpha^2}/{2M_\alpha})\ ,
\end{equation}         
which is to be understood as the two--body operator embedded in the
three--body space, with the relative kinetic energy operator ${\hat{\bf
q}_\alpha^2}/{2M_\alpha}$ of  particle $\alpha$ being subtracted from the
total energy variable $z$.  Considered in momentum space, Eq.(\ref{shift})
thus reads
\begin{equation}
\label{shiftmom}
    \langle{\bf p}'_\alpha,{\bf q}'_\alpha|T_\alpha(z)|
    {\bf p}_\alpha,{\bf q}_\alpha \rangle=
    \delta({\bf q}'_\alpha-{\bf q}_\alpha)
    \langle{\bf p}'_\alpha|t_\alpha(z-{q_\alpha^2}/{2M_\alpha})|
    {\bf p}_\alpha\rangle\ .
\end{equation} 
  
Since we are interested in $\eta d$ collision the subscript $\alpha$ is
fixed to 1.  Instead of the 9 equations of system (\ref{AGS}) we therefore
have to consider only the three equations
\begin{eqnarray}
\nonumber
        U_{11}(z)&=&\phantom{G_0^{-1}(z)+}\ \,
        T_2(z)G_0(z)U_{21}(z)+T_3(z)G_0(z)U_{31}(z)\ ,\\[0.2cm]
\label{AGSe}
        U_{21}(z)&=&G_0^{-1}(z)+
        T_1(z)G_0(z)U_{11}(z)+T_3(z)G_0(z)U_{31}(z)\ ,\\[0.2cm]
\nonumber
        U_{31}(z)&=&G_0^{-1}(z)+
        T_1(z)G_0(z)U_{11}(z)+T_2(z)G_0(z)U_{21}(z)\ ,
\end{eqnarray}
involving the operator $U_{11}$ which determines the elastic amplitude
(\ref{eampl}).  In momentum space the system (\ref{AGSe}) consists, after
partial wave decomposition, of three coupled two--dimensional integral
equations.  It is customary to reduce the dimension of these equations by
approximating or representing the two--body $T$--operators in separable form,
\begin{equation}
	t_\alpha(z)=|\chi_\alpha\rangle \tau_\alpha(z)\langle \chi_\alpha|\ ,
	       \qquad \alpha=1,2,3\ ,
\label{tsep}
\end{equation}
or, according to (\ref{shift}), by
\begin{equation}
\label{Ttau}
    T_\alpha(z)=|\chi_\alpha\rangle
    \tau_\alpha(z-{\hat{\bf q}_\alpha^2}/{2M_\alpha})\langle \chi_\alpha|\ ,
	       \qquad \alpha=1,2,3\ .
\end{equation}
Inserting this representation in the AGS equations (\ref{AGS}) and	       
sandwiching them between $\langle\chi_\beta|$ and $|\chi_\alpha\rangle$
they take the form
\begin{equation}
\label{Xab}
  X_{\beta\alpha}(z)=(1-\delta_{\beta\alpha})\langle\chi_\beta|
  G_0(z)|\chi_\alpha\rangle+
  \sum_{\gamma=1}^3(1-\delta_{\beta\gamma})\langle\chi_\beta|
   G_0(z)|\chi_\gamma\rangle \tau_\gamma(z-\frac{\hat{\bf          
   q}_\gamma^2}{2M_\gamma})X_{\gamma\alpha}(z)\ .
\end{equation}   
The three--body operators $U_{\beta\alpha}$, hence, are replaced 
by the effective two--body operators
\begin{equation}
\label{Xdef}
      X_{\beta\alpha}(z)=\langle\chi_\beta|
      G_0(z)U_{\beta\alpha}(z)G_0(z)|\chi_\alpha\rangle
\end{equation}
which act exclusively on the relative momentum states $|{\bf
q}_\alpha\rangle$. After partial wave decomposition we then end up with
one-dimensional integral equations.

The form-factor $|\chi_1\rangle$ in the $NN$ operator $t_1$ is related to the 
deuteron wave function $|\psi_d\rangle$ according to
\begin{equation}
\label{eass}
   |\psi_d;{\bf q}_1\rangle=
   G_0\left({q_1^2}/{2M_1}+E_d\right)|\chi_1\rangle|{\bf q}_1\rangle\ ,
\end{equation}
where $E_d$ is the deuteron energy. The on-energy-shell matrix element in
(\ref{eampl}), thus, can be written in the form
\begin{equation}
\label{eampl1}
	  f({\bf q}'_1,{\bf q}_1)=-(2\pi)^2M_1
	  \langle {\bf q}'_1|X_{11}(z)|{\bf q}_1\rangle\ .
\end{equation}
In other words, the set of equations (\ref{Xab}) provides the elastic
amplitude we are interested in.

In the present case there are further simplifications caused by the identity
of the nucleons. Indeed, the momentum representation of $X_{31}$, $\tau_3$,
$|\chi_3\rangle$ and of $X_{21}$, $\tau_2$, $|\chi_2\rangle$ respectively are
of the same functional form. This reduces the three equations involving
$X_{11}$ to the following pair
\begin{eqnarray}
\nonumber
	X_{11}(z) &=& 2\langle\chi_1|G_0(z)|\chi_2\rangle\tau_2\left(
	z-{\hat{\bf q}_2^2}/{2M_2}\right)X_{21}(z)\ ,\\[0.2cm]
\label{ident}
	X_{21}(z) &=& \langle\chi_2|G_0(z)|\chi_1\rangle+
	\langle\chi_2|G_0(z)|\chi_1\rangle\tau_1\left(
	z-{\hat{\bf q}_1^2}/{2M_1}\right)X_{11}(z)\\[0.2cm]
\nonumber
	&\phantom{=}&+\langle\chi_2|G_0(z)|\chi_2\rangle\tau_2\left(
	z-{\hat{\bf q}_2^2}/{2M_2}\right)X_{21}(z)\ .
\label{Xij}
\end{eqnarray}

The $S$--wave nucleon--nucleon separable potential is adopted from 
Ref.\cite{garcilazo} with its parameters slightly modified  to be consistent 
with more recent $NN$ data (see Ref.\cite{She98}). The $\eta$--nucleon 
$T$--matrix is taken in the form
\begin{equation}
\label{tetan}
%    t_{\eta N}(p',p;z)=
%    \frac{\lambda}{({p'}^2+\alpha^2)(z-E_0+i\Gamma/2)(p^2+\alpha^2)}
     t_{\eta N}(p',p;z)=({p'}^2+\alpha^2)^{-1}
    \frac{\lambda}{(z-E_0+i\Gamma/2)}(p^2+\alpha^2)^{-1}
\end{equation}
consisting of two vertex functions and the $S_{11}$-propagator in between
\cite{Rak96}.  It corresponds to the process $\eta N\to S_{11}\to \eta N$
which at low energies is dominant. The range parameter $\alpha=3.316$
fm$^{-1}$ was determined in Ref. \cite{bennh}, while $E_0$ and $\Gamma$ are
the parameters of the $S_{11}$ resonance \cite{PDG},
$$
E_0=1535\,{\rm MeV}-(m_N+m_\eta)\ ,\qquad \Gamma=150\,{\rm MeV}\ .
$$
The strength parameter $\lambda$ is chosen to reproduce the
$\eta$--nucleon scattering length $a_{\eta N}$,
\begin{equation}
\label{t000}
  \lambda=\frac{\alpha^4(E_0-i\Gamma/2)}{(2\pi)^2\mu_{\eta N}}a_{\eta N}\ .
\end{equation}
It is customary to use complex $a_{\eta N}$ the imaginary part of which
accounts for the flux losses into the $\pi N$ channel. The value of $a_{\eta
N}$ is not accurately known. Different analyses \cite{batinic} provided for
$a_{\eta N}$  values in the range
\begin{equation}
\label{interval}
0.27\ {\rm fm}\le{\rm Re\,}a_{\eta N}\le 0.98\ {\rm fm}\ ,\qquad
0.19\ {\rm fm}\le{\rm Im\,}a_{\eta N}\le 0.37\ {\rm fm}\ .
\end{equation}
Recently, however, most of the authors agreed that ${\rm Im\,}a_{\eta N}$ is
around 0.3\,fm.  But for ${\rm Re\,}a_{\eta N}$ the estimates are still very
different (compare, for example, Refs.  \cite{finn} and \cite{speth}).  We,
therefore, fixed ${\rm Im\,}a_{\eta N}$ to 0.3\,fm and did calculations for
several values of ${\rm Re\,}a_{\eta N}$ within the above interval.

We solved Eqs.  (\ref{ident}) for  $\eta d$ collision energies varying from
zero ($\eta d$-threshold, $z=E_d$) up to 22\,MeV.  As is well known (see, for
example, \cite{Bel}), the kernels of Eqs.(\ref{ident}), when expressed in
momentum representation, have  logarithmic singularities for $z > 0$.  These
singularities stem from the Green's functions which in their denominators
involve two terms $q_\alpha^2/2M_\alpha$ and $p_\alpha^2/2\mu_\alpha$
corresponding to the pair of Jacobi momenta.  The singularities appear after
angular integration, and their position depends on the values of $p_\alpha$
and $q_\alpha$ (the so-called moving singularities).  In the numerical
procedure, we handle it with the method suggested in Ref.  \cite{Soh71}.  The
main idea of this method consists in expanding the unknown solutions (in the
area covering the singular points) in certain polynomials and subsequent
analytic integration of the singular part of the kernels.

%%%%%%%%%%%%%%%%%%%%%%%%%%%%%%%%%%%%

\begin{figure}
\begin{center}
\unitlength=1mm
\begin{picture}(128,130)
\put(-20,0){\epsfig{file=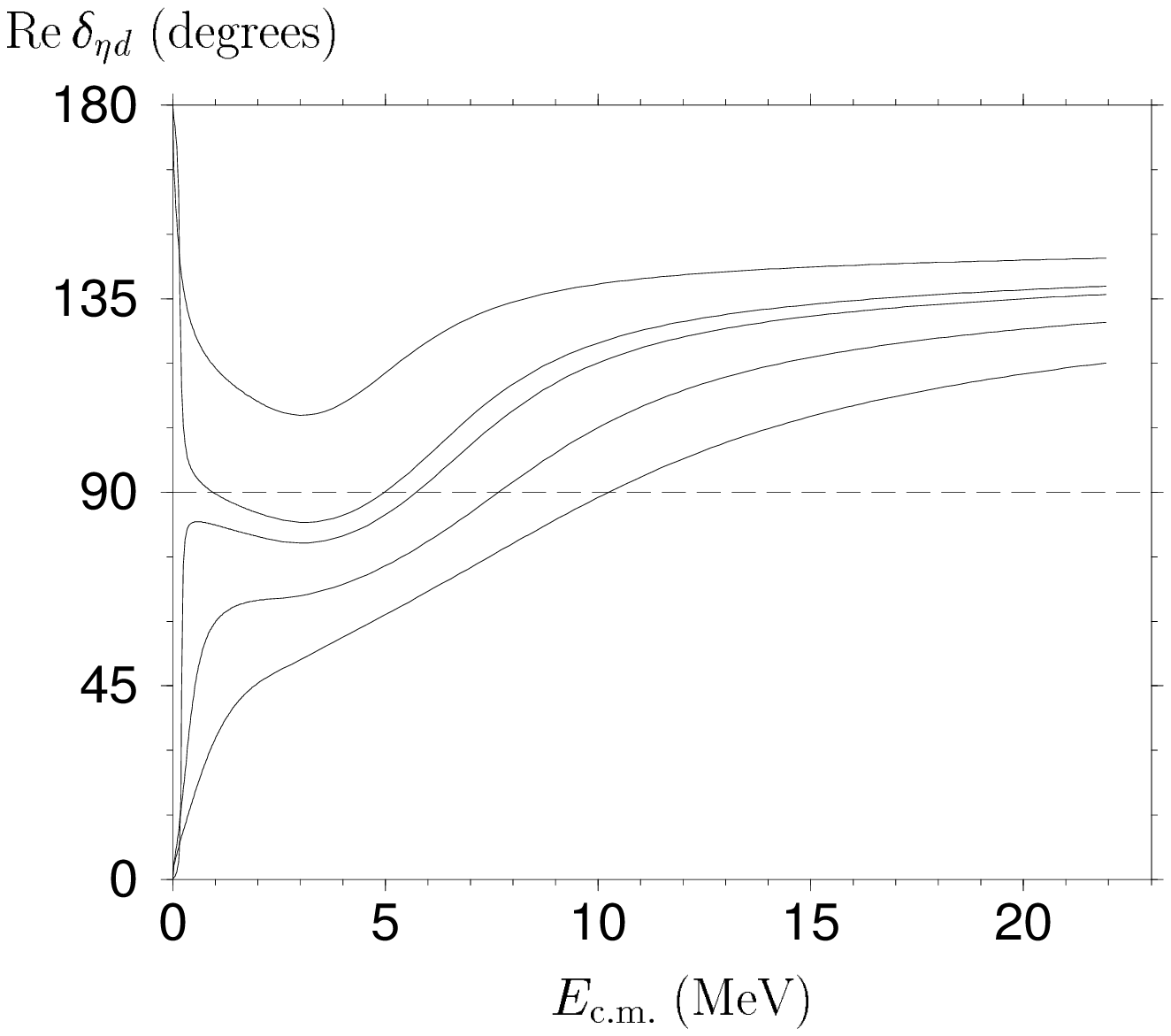}}
\end{picture}
\caption{
Real part of the $\eta$-deuteron phase-shift as a function of the collision
energy. The five curves correspond (starting from the lowest one) to
${\rm Re\,}a_{\eta N}$= 0.55\,fm, 0.65\,fm, 0.725\,fm, 0.75\,fm, and 0.85\,fm.
}
\label{red.fig}
\end{center}
\end{figure}

\begin{figure}
\begin{center}
\unitlength=1mm
\begin{picture}(128,150)
\put(-20,0){\epsfig{file=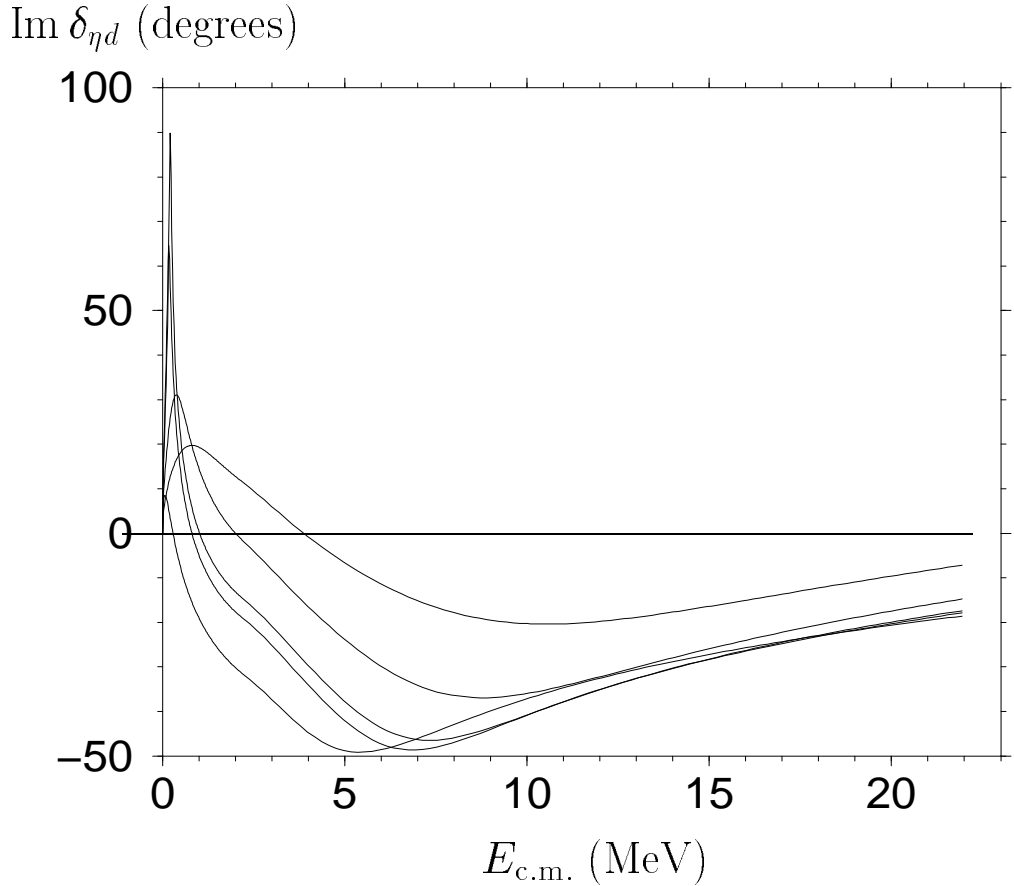}}
\end{picture}
\caption{
Imaginary part of the $\eta$-deuteron phase-shift as a function of the
collision energy.  The curves correspond to the same choices of the ${\rm
Re\,}a_{\eta N}$ as in Fig. 1. Within 1\,MeV to
6\,MeV, where the curves do not intersect, the lowest one corresponds to
${\rm Re\,}a_{\eta N}$=0.85\,fm, and every next curve above it corresponds to
a smaller ${\rm Re\,}a_{\eta N}$.
}
\label{imd.fig}
\end{center}
\end{figure}

\begin{figure}
\begin{center}
\unitlength=1mm
\begin{picture}(128,150)
\put(-20,0){\epsfig{file=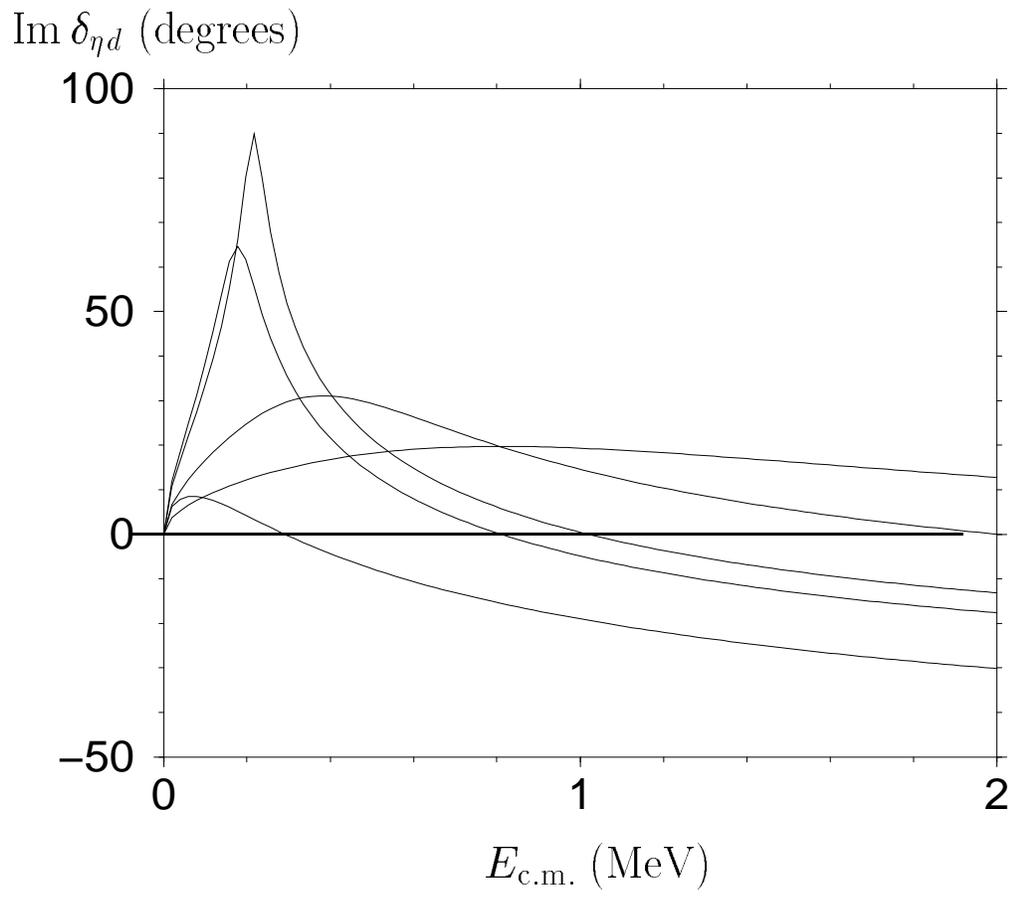}}
\end{picture}
\caption{
A magnified fragment of Fig. 2.
}
\label{im2d.fig}
\end{center}
\end{figure}

\begin{figure}
\begin{center}
\unitlength=1mm
\begin{picture}(128,150)
\put(-20,0){\epsfig{file=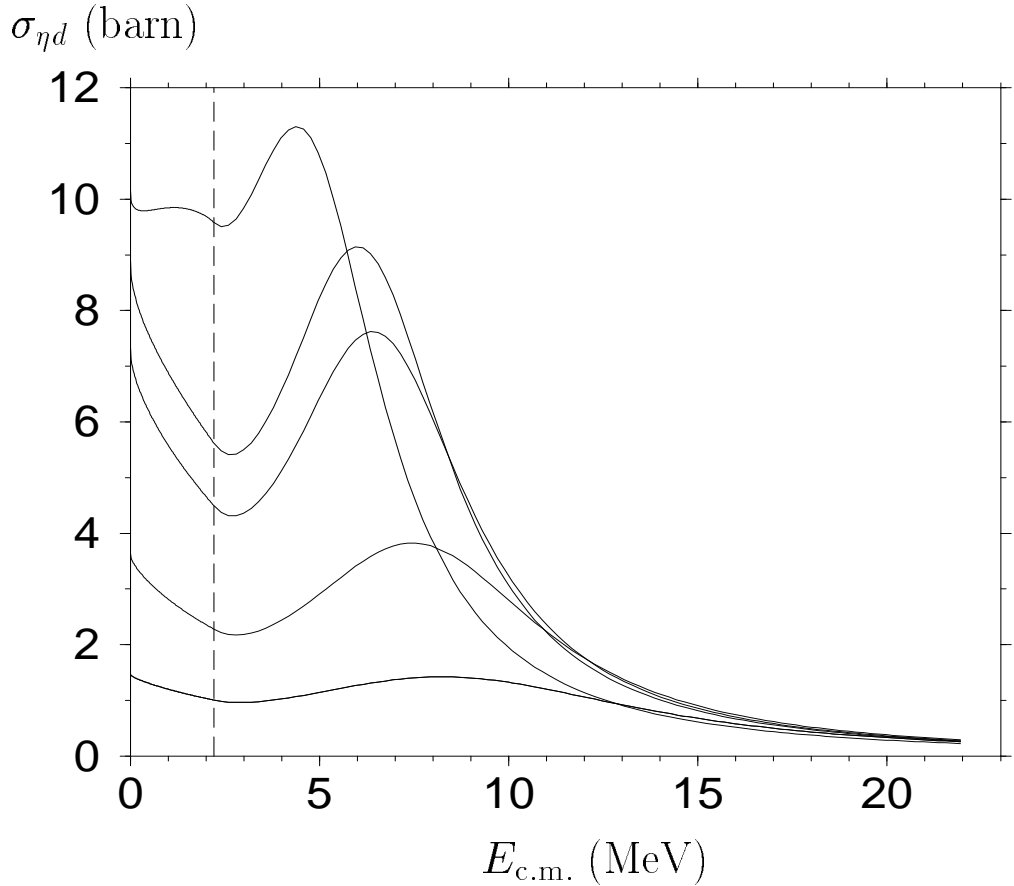}}
\end{picture}
\caption{
Total  cross--section  (integrated over the angles) for elastic
$\eta$--deuteron scattering as a function of collision energy. The five
curves correspond (starting from the lowest one) to ${\rm Re\,}a_{\eta N}$=
0.55\,fm, 0.65\,fm, 0.725\,fm, 0.75\,fm, and 0.85\,fm. The dashed line
indicates the deuteron break-up threshold.
}
\label{sig.fig}
\end{center}
\end{figure}

\begin{figure}
\begin{center}
\unitlength=1mm
\begin{picture}(128,130)
\put(-20,0){\epsfig{file=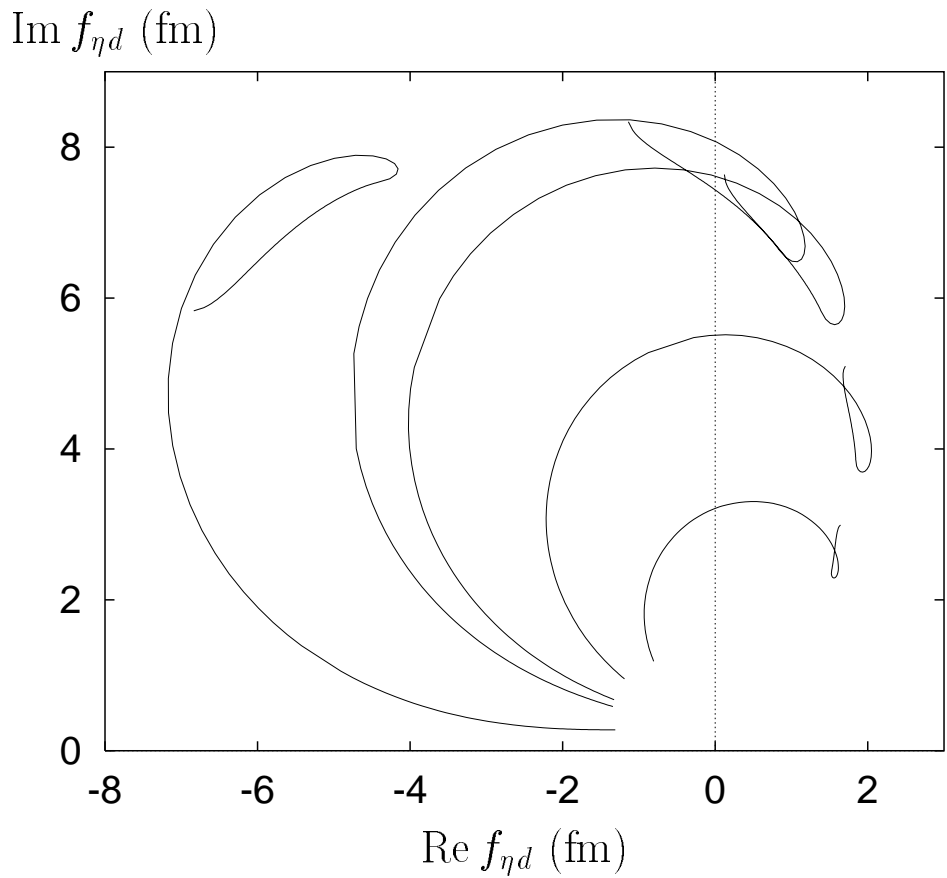}}
\end{picture}
\caption{
Argand plot for the $\eta d$ elastic scattering amplitude in the energy
interval from 0 to 22\,MeV. The five curves correspond (from right to left)
to ${\rm Re\,}a_{\eta N}$= 0.55\,fm, 0.65\,fm, 0.725\,fm, 0.75\,fm, and
0.85\,fm. When the energy increases the corresponding points move
anticlockwise.
}
\label{argand.fig}
\end{center}
\end{figure}

The results of our calculations are presented in
Figs.~\ref{red.fig}--\ref{argand.fig} and in Table~\ref{res.tab}.  In
Figs.~\ref{red.fig} and \ref{imd.fig} the energy dependence of the $\eta d$
phase-shifts for five different choices of ${\rm Re\,}a_{\eta N}$ is shown,
namely, for 0.55\,fm, 0.65\,fm, 0.725\,fm, 0.75\,fm, and 0.85\,fm.  The
larger this value, the stronger is the $\eta$N attraction.  The change in the
character of these curves, hence, reflects the growth of the attractive force
between the $\eta$ meson and the nucleon.  The lower three curves for ${\rm
Re\,}\delta_{\eta d}$ corresponding to the smaller values of ${\rm
Re\,}a_{\eta N}$ start from zero, the two curves corresponding to the strong
attraction start from $\pi$.  According to Levinson's theorem, the phase
shift at threshold energy is equal to the number of bound states $n$ times
$\pi$.  We found that the transition from the lower family of the curves to
the upper one happens at ${\rm Re\,}a_{\eta N}$=0.733\,fm.  Therefore, the
$\eta$N force, which generates ${\rm Re\,}a_{\eta N}>0.733$\,fm, is
sufficiently attractive to bind $\eta$ inside the deuteron.  This is the
first conclusion of our calculations.

The second conclusion concerns the peaks in the energy dependence of the
total elastic cross-section (see Fig.~\ref{sig.fig}), indicating that a
resonance appears in the $\eta d$--system. Of course, not every maximum of
the cross-section is a resonance, but the Argand plots, shown in
Fig.~\ref{argand.fig}, prove that the maxima we found are resonances.  Their
positions and widths for various choices of ${\rm Re\,}a_{\eta N}$ are given
in Table \ref{res.tab}.  It should be noted that, while the resonance energy
is determined in our calculations exactly (as the maximum of the function
$\sin^2{\rm Re\,}\delta_{\eta d}$), the corresponding width is obtained by
fitting the cross-section with a Breit--Wigner curve.  Therefore, the values
of $\Gamma_{\eta d}$ given in Table \ref{res.tab} should be considered only
as rough estimates.

The presence of a resonance before a quasi-bound state appears is not
surprising.  With increasing attraction the poles of the $S$-matrix should
move in the complex plane from the resonance area to the quasi-bound state
area. This is exactly what our calculations indicate.  In Ref.\cite{She98} we
showed that such transition of the pole happens when ${\rm Re\,}a_{\eta N}$
changes from 0.25\,fm to 1\,fm.  Here we found more exactly that it happens
at ${\rm Re\,}a_{\eta N}$=0.733\,fm.

The resonant behaviour of $\eta d$ elastic scattering should be seen in
various processes involving $\eta d$--system in their final states, such as
$\gamma d\to d\eta$ and $np\to d\eta$.  Recent measurements \cite{calen} of
the $\eta$--production in the $np$ collisions reveal a bump of the
cross--section for the reaction $np\to d\eta$ at a c.m.  energy below 5\,MeV.
If we suppose that the energy dependence of this cross--section is mainly
determined by the final state interaction, then this bump can be explained by
the existence of an $\eta d$ resonance.  Moreover, since the bump was
observed below 5\,MeV, the resonance positions given in Table~\ref{res.tab}
imply the rough lower bound ${\rm Re\,}a_{\eta N}\ge 0.75\,{\rm fm}$.  For a
reliable estimate, however, one has to perform an explicit calculation of the
corresponding cross-section.  A recent analysis \cite{speth} of this reaction
with a non-resonant final state interaction is consistent with the data of
Ref.\cite{calen} only if ${\rm Re\,}a_{\eta N}=0.30\,{\rm fm}$. Finally, it
is interesting to note that there is tentative evidence for a similar bump in
the low--energy region in the reaction $\gamma d\to X\eta$
\cite{kruschemetag}.

%%%%%%%%%%%%%%%%%%%%%
%%%%%%%%%%%%%%%%%%%%%%%%%%%%%%%%%%%%%%%%%%%%%%%%
%%%%%%%%%%%%%%%      ACKNOWLEDGEMENTS  %%%%%%%%%
%%%%%%%%%%%%%%%%%%%%%%%%%%%%%%%%%%%%%%%%%%%%%%%%

\bigskip
\acknowledgements{
Financial support from Deutsche Forschungsgemeinschaft, NATO (grant 
\#CRGLG970110), INTAS (grant \#96-0457), International Center for
Fundamental Physics (Moscow), Russian Foundation for Basic Research, National
Research Foundation of South Africa and the Joint Institute  for Nuclear
Research (Dubna) is greatly appreciated. }

%%%%%%%%%%%%%%%%%%%%%%%%%%%%%%%%%%%%%%%%%%%%%%%%%%%%%%%%%%%%
%%%%%%%%%%%        REFERENCES   %%%%%%%%%%%%%%%%%%%%%%%%%%%%
%%%%%%%%%%%%%%%%%%%%%%%%%%%%%%%%%%%%%%%%%%%%%%%%%%%%%%%%%%%%

%%%%%%%%%%%%%%%%%%%%%%%%%%%%%%%%%%%%%%%%%%%%%%%%%%%%%%%%
%%%%%%%%%%%%%%%    FIGURES %%%%%%%%%%%%%%%%%%%%%%%%%%%%%
%%%%%%%%%%%%%%%%%%%%%%%%%%%%%%%%%%%%%%%%%%%%%%%%%%%%%%%%

%%%%%%%%%%%%%%%%%%%%%%%%%%%%%%%%%%%%%%%%%%%%%%%%%%%%%%%%%%%%
%%%%%%%%%%%%%%%%%%%%%%%%%%%%%%%%%%%%%%%%%%%%%%%%%%%%%%%%%%%%
%%%%%%%%%%%%%%%%%%%%%%%%%%%%%%%%%%%%%%%%%%%%%%%%%%%%%%%%%%%%
%%%%%%%%%%%%%%%%%%%%%%%%%%%%%%%%%%%%%%%%%%%%%%%%%%%%%%%%%%%%
%%%%%%%%%%%%%%%%%%%%%%%%%%%%%%%%%%%%%%%%%%%%%%%%%%%%%%%%%%%%%%%%%%%%%%%%
%                    T A B L E
%%%%%%%%%%%%%%%%%%%%%%%%%%%%%%%%%%%%%%%%%%%%%%%%%%%%%%%%%%%%%%%%%%%%%%%%
\begin{table}
\begin{tabular}{|c|c|c|}
\hline
 ${\rm Re\,}a_{\eta N}$ (fm) &    $E_{\eta d}^{\rm res}$ (MeV) &
 $\Gamma_{\eta d}$ (MeV)\\
\hline
   0.55          &             8.24        &        9.15\\
   0.65          &             7.46        &        8.45 \\
   0.675         &             7.14        &        7.61  \\
   0.70          &             6.79        &        6.90   \\
   0.725         &             6.41        &        6.31    \\
   0.75          &             6.01        &        5.87     \\
   0.85          &             4.39        &        5.79      \\
   0.90          &             3.73        &        6.81       \\
\hline
\end{tabular}                    
\caption{
Energy and width of the $\eta d$ resonance for various choices of
${\rm Re\,}a_{\eta N}$.
}
\label{res.tab}
\end{table}
\end{document}